\documentclass[12pt]{article}
\usepackage{amsfonts,amsmath,amssymb,epsf,cite}
\usepackage{graphicx,color,subfigure}
\usepackage[usenames,dvipsnames]{pstricks}
 \def\ben{\begin{equation}}
\def\een{\end{equation}}
\def\beq{\begin{eqnarray}}
\def\eeq{\end{eqnarray}}
\def\bea{\begin{eqnarray}}
\def\eea{\end{eqnarray}}

\def\vx{{\vec{x}}}

\def\vy{{\vec{y}}}

\def\vk{{\vec{k}}}

\def\nn{\nonumber}
\def\vphi{\vec{\phi}}
\begin{document}
\begin{titlepage}
\thispagestyle{empty}
\begin{flushright}
UK/12-04\\
BROWN-HET-1635\\
\end{flushright}

\bigskip

\begin{center}
\noindent{\Large \textbf
{Bi-local Construction of $Sp(2N)$/dS Higher Spin  Correspondence}}\\
\vspace{2cm} \noindent{
Diptarka Das$^{(a)}$\footnote{e-mail:diptarka.das@uky.edu},
Sumit R. Das $^{(a)}$ \footnote{e-mail:das@pa.uky.edu},
Antal Jevicki $^{(b)}$ \footnote{e-mail: antal\_jevicki@brown.edu}
and Qibin Ye $^{(b)}$ \footnote{e-mail: qibin\_ye@brown.edu}\\
\vspace{1cm}
$^{(a)}$  {\it Department of Physics and Astronomy, \\
 University of Kentucky, Lexington, KY 40506, USA}\\
\vspace{0.5cm}
$^{(b)}$ {\it Department of Physics,\\
Brown University, Providence, RI 02912, USA}}
\end{center}

\vspace{0.3cm}
\begin{abstract}
We derive a collective field theory of the singlet sector of the $Sp(2N)$
sigma model. Interestingly the Hamiltonian for the bilocal collective
field is the {\em same} as that of the $O(N)$ model. However, the
large-$N$ saddle points of the two models differ by a sign. This leads
to a fluctuation Hamiltonian with a negative quadratic term and
alternating signs in the nonlinear terms which correctly reproduces the correlation functions of 
the singlet sector. Assuming the validity of the
connection between $O(N)$ collective fields and higher spin fields in
AdS, we argue that a natural interpretation of this theory is by a
double analytic continuation, leading to the dS/CFT correspondence
proposed by Anninos, Hartman and Strominger. The bi-local construction gives a map into the bulk of 
de Sitter space-time.   Its geometric pseudospin-representation provides a framework for quantization and definition of the Hilbert space.
We argue that this is consistent with  finite $N$ Grassmannian constraints, establishing the bi-local representation as
 a nonperturbative framework for quantization of Higher Spin Gravity in de Sitter space.

\end{abstract}
\end{titlepage}
\newpage

\tableofcontents
\newpage

\section{Introduction and summary}

The proposed duality \cite{Klebanov:2002ja} of the singlet sector of
the $O(N)$ vector model in three space-time dimensions and Vasiliev's
higher spin gauge theory in AdS$_4$ \cite{Vasiliev:2012vf} has received a definite verification\cite{Giombi:2009wh,Sezgin:2003pt}
and has also thrown  valuable light on the origins of holography. Since the field theory is
solvable in the large-N limit, one might hope that there is an
explicit derivation of the higher spin gauge theory from the vector
model, thus providing an explicit understanding of the emergence of
the holographic direction. Indeed, the singlet sector of the $O(N)$
model can be expressed in terms of a Hamiltonian for the bi-local
collective field, $\sigma(\vx,\vy) = \phi^i (\vx) \phi^i (\vy)$
where $\phi^i(\vx), i= 1 \cdots N$ is the $O(N)$ vector field. In \cite{Das:2003vw} it was proposed that
Vasiliev's fields are in fact components of $\sigma(\vx,\vy)$. The
precise connection between the bi-local and HS bulk fields was written explicitly  in the 
light cone frame\cite{Koch:2010cy, Jevicki:2011aa, Jevicki:2011ss}: 
the correspondence in general  involves a nonlocal transformation corresponding to
a canonical transformation in phase space\footnote{See also reference \cite{Douglas:2010rc}.}. This provides
a direct understanding of the emergence of a holographic direction
from the large-N degrees of freedom, in a way similar to the well
known example of the $c=1$ Matrix model \cite{Das:1990kaa}. In both
these models, the large-N degrees of freedom gave rise to an additional
dimension which had to be interpreted as a {\em spatial} dimension \footnote{Other instances 
of emergence of dimensions from large-N degrees of freedom, e.g. Eguchi-Kawai 
models \cite{Eguchi:1982nm}, Matrix Theory \cite{Banks:1996vh,Ishibashi:1996xs} 
also lead to spatial directions in Lorentzian signature or Euclidean theories.}.

In contrast to AdS/CFT correspondence, any dS/CFT correspondence \cite{dS}
involves an emergent holographic direction which is {\em timelike}. It is then of interest
to understand how a {\em timelike} dimension is generated from large-N degrees of freedom. Recently, 
Anninos, Hartman and Strominger \cite{Anninos:2011ui} put forward a conjecture that the {\em euclidean} $Sp(2N)$ vector
model in three dimensions is dual to Vasiliev higher spin theory in
four dimensional de Sitter space.  

In this work we construct a collective field theory of the {\em Lorentzian} $Sp(2N)$ model which captures the singlet state dynamics of
the $Sp(2N)$ vector model. Using the results of \cite{Das:2003vw} and
\cite{Koch:2010cy} we then argue that a natural interpretation of the
resulting action is by double analytic continuation which makes the emergent direction time-like, relating this to  higher spin theory in dS$_4$, in a way
reminiscent of the way the Louiville mode in worldsheet string theory has to be interpreted as a time beyond critical 
dimensions \cite{Das:1988ds}. Our map establishes the bi-local theory as the bulk space-time  representation of de Sitter higher spin gravity.

The bilocal collective field is a composite of two Grassmann variables and therefore might not appear to be a genuine bosonic field. In particular  for finite $N$ a sufficiently large power of the field operator vanishes, reflecting its Grassmannian origin \footnote{This property of higher spin currents has been already recognized in \cite{Ng:2012xp}}. This is further reflected on the size of its Hilbert space. The bulk theory cannot be a usual bosonic theory defined on dS space, though it may be regarded as such in a perturbative $1/N$ expansion.

The implementation of the Grassmann origin of the Hilbert space will be given a central attention  in the present work. For this we will describe a geometric (pseudo-spin) version of the collective theory  which will be seen to incorporate these effects.  For dS/CFT, this implies that the true number of degrees of freedom in the dual higher spin theory in dS is in this framework reduced from what is  seen perturbatively (with $G=R_{dS}^2/N$ being the coupling constant squared).
The issue of the size of the Hilbert space is of central relevance for possible accounting of entropy of de Sitter space. 
For pure Gravity in de Sitter space, it has argued that the Entropy being $S=A/4G$  with a finite area of the horizon requires a finite dimensional Hilbert space \cite{witten,Balasubramanian:2001rb,Banks:2003cg}. Interesting quantum mechanical models have been  proposed \cite{Balasubramanian:2001rb,Volovich:2001rt,Parikh:2004wh,Lowe:2004nw}  to account for this. But apparent conflicts  between a finite entropy of de Sitter space with the usual formulations of dS/CFT have been discussed for example in in \cite{Dyson:2002nt}. In the present case of dS/CFT we are dealing with N-component quantum field theory with d=3 dimensional space so clearly the number of degrees of freedom must be infinite. Consequently the question of Entropy remains open and is an interesting  topic for further investigations.

\section{The $Sp(2N)$ vector model}

The $Sp(2N)$ vector model in $d$ spacetime dimensions is defined by the action
\ben
S = i\int dt d^{d-1}x~\left[ \{ \partial_t \phi^i_1\partial_t
  \phi^i_2 - \nabla  \phi^i_1 \nabla \phi^i_2\} -V(i\phi^i_1\phi^i_2)
  \right]
\label{one}
\een
where $\phi^i_1,\phi^i_2$ with $i=1\cdots N$ are $N$ pairs of
Grassmann fields.  This is of course a model of ghosts. 

In this section we will quantize this model following \cite{Henneaux:1982ma} and \cite{Finkelstein:1985cx}. 
In this quantization, the fields $\phi_1^i$ and $\phi_2^i$ are hermitian operators, while the canonically conjugate momenta
\ben
P_1^i = i\partial_t \phi^i_2~~,~~~~~~~~~P_2^i = - i\partial_t \phi^i_1
\label{two}
\een
are anti-hermitian. The Hamiltonian $H$ is hermitian
\ben
H =i \int d^{d-1}x \left[ P^i_2 P^i_1 + \nabla \phi^i_1 \nabla \phi^i_2 + V(i\phi^i_1\phi^i_2) \right]
\label{three}
\een
The (equal time) canonical anticommutation relations are
\bea
\{ \phi^a_i (\vx) , P^b_j (\vy) \} & = & -i \delta_{ij}\delta^{ab}\delta^{d-1}(\vx - \vx^\prime) \nn \\
\{ \phi^i_a (\vx) , \phi^j_b (\vy) \} & = & \{ P^i_a (\vx) , P^j_b (\vy) \} = 0~~,~~~~~~~~~~(a,b=1,2)
\label{four}
\eea
with all other anticommutators vanishing.  With these anticommutators the equations of motion for the corresponding Heisenberg picture operators
\ben
\partial_t^2 \phi^i_a - \nabla^2 \phi^i_a + V^\prime = 0
\label{five}
\een
follow. 
The operator relations (\ref{four}) allow a representation of the operators as follows
\ben
\phi^a_i (\vx) \rightarrow \phi^a_i (\vx)~~,~~~~~~P^a_i \rightarrow -i \frac{\delta}{\delta \phi^a_i(\vx)}
\label{six}
\een
where $\phi^i_a$ are now Grassmann fields.

For the free theory, the solution to the equation of motion is
\ben
\phi^i_a (\vx,t) = \int\frac{d^{d-1}k}{(2\pi)^{d-1}\sqrt{2|k|}} \left[
  \alpha_a^i (\vk) e^{-i(|k|t - \vk \cdot \vx)} + \alpha_a^{i\dagger}(\vk)e^{i(|k|t - \vk
    \cdot \vx)} \right]
\label{seven}
\een
and the operators $\alpha^i_a$ satisfy 
\ben \{ \alpha_1^i (\vk), \alpha_2^{\dagger j} (\vk^\prime) \} = i
\delta^{ij} \delta (\vk - \vk^\prime) ~~,~~~~~\{ \alpha_1^{\dagger
  i}(\vk), \alpha_2^j (\vk^\prime) \} = -i \delta^{ij}\delta (\vk -
\vk^\prime)
\label{eight}
\een
with all the other anticommutators vanishing. 
The Hamiltonian is given by
\ben
H = i\int [d\vk]~|\vk| \left[\alpha_1(\vk)^\dagger \alpha_2 (\vk) - \alpha_2 (\vk)^\dagger \alpha_1 (\vk) \right]
\een
The basic commutators lead to 
\ben
[ H , \alpha^i_a (k)] = - k \alpha^i_a (k)~~,~~~~
[ H , \alpha^{i\dagger}_a] = k \alpha^{i\dagger}_a (k)
\label{aone}
\een
To discuss the quantization of the free theory it is useful to review
the quantization of the $Sp(2N)$ oscillator, 
following \cite{Finkelstein:1985cx} \footnote{Note that our
  notation is different from that of \cite{Finkelstein:1985cx}}. The Hamiltonian is
\ben
H = i(-\frac{\partial^2}{\partial \phi_2^i \partial \phi_1^i}+k^2 \phi^i_1 \phi^i_2)
\label{four-one}
\een
where $\phi^i_1,\phi^i_2$ are $N$ pairs of Grassmann numbers. 
Because of the Grassmann nature of the variables the
spectrum of the theory is bounded both from below and from above. 
The oscillators are defined by (in the Schrodinger picture) \ben \phi_a^i =
\frac{1}{\sqrt{2k}} [ \alpha_a^i + \alpha_a^{i\dagger} ]
\label{asix}
\een
while the momenta are
\ben
P_a^i = \epsilon_{ab} \sqrt{\frac{k}{2}}(\alpha_b^i - \alpha_b^{i\dagger})
\label{aseven}
\een
The ground state $|0\rangle$ and the highest state $|2N\rangle$ are then given by the conditions
\ben
\alpha^i_a |0\rangle = 0
~~,~~~~~~~~~~~~~~~\alpha^{i\dagger}_a |2N\rangle = 0
\label{atwo}
\een
with the wavefunctions
\ben
\Psi_0 = {\rm exp}[-ik \phi_1^i \phi_2^i]~~,~~~~~~~~~~~~~
\Psi_{2N} = {\rm exp}[ik \phi_1^i \phi_2^i]
\label{athree}
\een
and the energy spectrum is given by
\ben
E_n = k [n - N]~~,~~~~~~~~~~~~~~n = 0,1,\cdots, 2N
\label{afour}
\een
Finally, the Feynman correlator of the Grassmann coordinates may be easily seen
to be
\ben
\langle 0|T[\phi^i_1(t)\phi^j_2(t^\prime)]|0\rangle= \frac{i\delta^{ij}}{2k}~e^{-ik|t-t^\prime|}
\label{afive}
\een
Extension of these results to the free field theory is straight forward: for each momentum $\vk$, we have a fock space with a finite number of states. 

\section{Collective Field Theory for the $Sp(2N)$ model}

In the representation (\ref{six}) a general wavefunctional is given by $\Psi[\phi^i_a(\vx),t]$. Our aim is to obtain a description of the singlet sector of the theory, i.e. wavefunctionals which are invariant under the $Sp(2N)$ rotations of the fields $\phi^i_a(\vx)$. All the invariants in field space are functions of the bilocal collective fields
\ben
\rho(\vx,\vy) \equiv i \epsilon^{ab}\phi^i_a (\vx) \phi^i_b (\vy)
\label{twelve}
\een

We have defined this collective field to be hermitian (which is why there is a $i$ in the definition). Clearly $\rho(\vx,\vy) = \rho(\vy,\vx)$. The aim now is to rewrite the theory in terms of a Hamiltonian which is a functional of $\rho (\vx,\vy)$ and its canonical conjugate $-i \frac{\delta}{\delta \rho(\vx,\vy)}$ which acts on wavefunctionals which are functionals of $\rho (\vx,\vy)$.

It is important to remember that $\rho(\vx,\vy)$ is not a genuine bosonic field. This will have important consequences at finite $N$. In a perturbative expansion in $1/N$, however,
there is no problem \cite{deMelloKoch:1996mj} in  treating $\rho(\vx,\vy)$ as a bosonic field.

Before dealing with the $Sp(2N)$ field theory, it is useful to review some aspects of the collective theory for the usual $O(N)$ model, starting with the $O(N)$ oscillator.

\subsection{Collective fields for the $O(N)$  theory}

In this section we review the bi-local collective field theory construction for the $O(N)$ field theory, starting with the $O(N)$ oscillator. This has a Hamiltonian
\ben
H = \frac{1}{2} [ P^iP^i + k^2 X^i X^i]
\label{fourteen}
\een
The collective variable is the square of the radial coordinate 
$\sigma = X^i X^i$ and the Jacobian for transformation from $X^i$ to $\sigma$ and the angles is
\ben
J(\sigma) = \frac{1}{2}t  \sigma^{(N-2)/2}\Omega_{N-1} 
\label{eighteen}
\een
where $\Omega_{N-1}$ is the volume of unit $S^{N-1}$.
The idea is to find the Hamiltonain $H(\sigma, \frac{\partial}{\partial \sigma})$ which acts on wavefunctions $[J(\sigma)]^{1/2} \Psi (\sigma)$.
The key observation of \cite{Jevicki:1979mb} is that this can also be obtained by requiring that $H(\sigma, \frac{\partial}{\partial \sigma})$ acting on wavefunctions $[J(\sigma)]^{1/2} \Psi (\sigma)$ is hermitian with the trivial measure $d\sigma$. This determines both the Jacobian and the Hamiltonian and the technique generalizes to higher dimensional field theory. The final result is well known,
\ben
H_{coll} = -2\frac{\partial}{\partial \sigma} \sigma \frac{\partial}{\partial \sigma} + \frac{(N-2)^2}{8\sigma} + \frac{1}{2} k^2 \sigma
\label{twothree}
\een
The large-$N$ expansion then proceeds as usual by expanding around the saddle point solution $\sigma_0$ which minimizes the potential \footnote{To see why the saddle point approximation is valid, rescale 
$\sigma \rightarrow N\sigma$ and $\Pi_\sigma \rightarrow \frac{1}{N} \Pi_\sigma$ so that there is an overall factor of $N$ in front of the potential energy term. We will, however, stick to the unrescaled fields.},
\ben
\sigma_0^2 = \frac{N^2}{4k^2}
\label{twofive}
\een
Clearly, we have to choose the positive sign since in this case $\sigma$ is a {\em positive} real quantity,
\ben
\sigma_0 = \frac{N}{2k}
\label{twosix}
\een
which reproduces the coincident time two point function $\langle 0|X^i(t)X^i(t)|0\rangle$ and the correct ground state energy, $
E_0 = \frac{N}{2}k $. The subleading contributions are then obtained by expanding around the saddle point, 
\ben
\sigma = \sigma_0 + \sqrt{\frac{2N}{k}}\eta~~,~~~~~\Pi_\sigma =
\sqrt{\frac{k}{2N}}
\pi_\eta
\label{twoeight}
\een
The quadratic part of the Hamiltonian becomes
\ben
H^{(2)} = \frac{1}{2} \left[\pi_\eta^2+4k^2 \eta^2 \right]
\label{bone}
\een
This leads to the excitation spectrum to $O(1)$,
$E_n = 2n k$ with $n = 0,1,\cdots ,\infty$.
The Hamiltonian of course contains all powers of $\eta$.  {\em 
Terms with even number of the fluctuations $(\pi_\eta,\eta)$ come with odd factors of $\sigma_0$. } This fact will play a key role in the following.

In the following it will be necessary to consider wavefunctions. It follows directly from (\ref{fourteen}) that the ground state wavefunction is given by (up to a normalization which is not important for our purposes)
\ben
\Psi_0 (X^i) = {\rm exp} [ -\frac{k}{2} \sigma ] \sim {\rm exp} [ -\sqrt{\frac{Nk}{2}} \eta ]
\label{bthree}
\een
where we have expanded $\sigma$ as in (\ref{twoeight}), used (\ref{twosix}) and ignored an overall constant.
We should get the same result from the collective theory. Recalling that the collective wavefunction is related to the original wavefunction by a Jacobian factor, the ground state wavefunction follows from (\ref{bone})
\ben
\Psi_0^\prime(\eta) = [J(\sigma)]^{-\frac{1}{2}}~{\rm exp} [-k \eta^2]
\label{bfive}
\een
The presence of the Jacobian is crucial in obtaining agreement with (\ref{bthree}) \cite{Jackiw:1980xj}.
Expanding the argument in the Jacobian in powers of $\eta$ according to (\ref{twoeight}) it is easy to see that the quadratic term in $\eta$ coming from the Jacobian exactly cancels the explicit quadratic term in (\ref{bfive}) and the linear term in $\eta$ is in exact agreement with (\ref{bthree}). The expression (\ref{bfive}) of course contain all powers of $\eta$ once exponentiated - these should also cancel once one takes into account the cubic and higher terms in the collective Hamiltonian as well as finite $N$ corrections which we have ignored to begin with. 
The above formalism can be easily generalized to an additional invariant potential, since the latter would be a function of $\sigma$.

The collective theory for $O(N)$ field theory can be constructed along identical lines. We reproduce the relevant formulae
from \cite{Jevicki:1979mb} which are direct generalizations of
the formulae for the oscillator. The $O(N)$ model has the Hamiltonian
\ben 
H = \frac{1}{2} \int d^{d-1}x \left[ -\frac{\delta^2}{\delta
    \phi^i(\vx) \delta \phi^i(\vx)}+ \nabla \phi^i (\vx) \nabla \phi^i
  (\vx) + U[\phi^i(\vx) \phi^i(\vx)] \right]
\label{twonine}
\een
The singlet sector Hamiltonian in terms of the bi-local collective field
$\sigma(\vx,\vy) = \phi^i(\vx)\phi^i(\vy)$ and its canonically conjugate momentum $\Pi_\sigma(\vx,\vy)$ is, to leading order in $1/N$ \footnote{To subleading order there are singular terms which are crucial for reproducing the correct $1/N$ contributions.}
\ben
H^{O(N)}_{coll} = 2 {\rm{Tr}} \left[ (\Pi_\sigma \sigma \Pi_\sigma) +\frac{N^2}{16} \sigma^{-1} \right] - \frac{1}{2}\int d\vx \nabla_x^2 \sigma (\vx,\vy) |_{\vy = \vx} + U(\sigma(\vx,\vx))
\label{three-three}
\een
where  the spatial coordinates are treated as matrix indices. 

So far our considerations are valid for an arbitrary interaction potential $U$. Let us now restrict ourselves to the free theory, $U = 0$ to discuss the large-$N$ solution explicitly. 
In momentum space
the saddle point solution is
\ben
\sigma(\vk_1,\vk_2) = \frac{N}{2|\vk_1|}\delta (\vk_1-\vk_2)
\een
Once again we have chosen the positive sign in the solution of the saddle point equation, and the saddle point value of the collective field agrees with the two point correlation function of the basic vector field, which should be positive. The $1/N$ expansion is generated in a fashion identical to the single oscillator,
\ben
\sigma(\vk_1,\vk_2) = \sigma_0 (\vk_1,\vk_2)+ \left( \frac{|\vk_1||\vk_2|}{N(|\vk_1|+|\vk_2|)} \right)^{-\frac{1}{2}}\eta(\vk_1,\vk_2)~~,~~\Pi_\sigma = \left( \frac{|\vk_1||\vk_2|}{N(|\vk_1|+|\vk_2|)} \right)^{\frac{1}{2}}\pi_\eta (\vk_1,\vk_2)
\label{three-five}
\een
the quadratic piece becomes
\ben
H^{(2)} = \frac{1}{2}\int d\vk_1 d\vk_2 \left[ \pi_\eta (\vk_1,\vk_2) \pi_\eta (\vk_1,\vk_2)+ (|\vk_1|+|\vk_2|)^2 \eta (\vk_1,\vk_2)\eta (\vk_1,\vk_2) \right]
\label{three-eight}
\een
so that the energy spectrum is given by
\ben
E(\vk_1, \vk_2) = |\vk_1|+|\vk_2|
\label{three-nine}
\een
as it should be. It is easy to check that the unequal time two point function of the fluctuations reproduces the connected part of the two point function of the full collective field as calculated from the free field theory.
A nontrivial $U$ can be reinstated easily (see e.g. the treatment of the $(\vphi^2)^2$ model in \cite{Das:2003vw}, which discusses the RG flow to the nontrivial IR fixed point).

\subsection{ Collective theory for the $Sp(2N)$ oscillator}

Since there is a representation of the field operator and the
conjugate momentum operator of the $Sp(2N)$ theory in terms of
Grassmann fields, (\ref{six}), it is clear that the derivation of the
collective field theory of the $Sp(2N)$ model closely parallels that of
the $O(N)$ theory. In this subsection we consider the $Sp(2N)$
oscillator. The Hamiltonian is given by (\ref{four-one}.
The collective variable is 
\ben
\rho = i \epsilon^{ab}\phi^i_a\phi^i_b
\label{four-two}
\een
The fully connected correlators
of this collective variable have a simple relationship with those of
the $O(2N)$ harmonic oscillator, 
\ben
\langle \rho(t_1)\rho(t_2)\cdots\rho(t_n)\rangle^{conn}_{Sp(2N)} = -
\langle \sigma(t_1)\sigma(t_2) \cdots \sigma(t_n)\rangle^{conn}_{SO(2N)}
\label{four-two1}
\een
This result follows from (\ref{afive}) and the application of Wick's
theorem for Grassmann variables.

The collective variable $\rho$ is a Grassmann even variable - it is not an usual bosonic variable. This key fact is intimately related to the finite number of states of the $Sp(2N)$ oscillator. In this section we will show that in a $1/N$ expansion we can nevertheless proceed, defering a proper discussion of this point to a later section.

The Hamiltonian for the collective theory is obtained by the same
method used to obtain the collective theory in the bosonic case, with various negative sign coming from the Grassmann nature of the variables.
Using the chain rule and taking care of negative signs coming because of Grassmann numbers, one gets the Jacobian
$J^\prime(\rho)$ (determined by requiring the hermicity of $J^{-1/2} H J^{1/2}$) 
\ben J^\prime(\rho) = A^\prime~\rho^{-(N+1)}
\label{four-five}
\een
where $A^\prime$ is a constant.
The negative power of $\rho$ of course reflects the Grassmann nature
of the variables \footnote{This $\rho$ dependence of the Jacobian
follows from a direct calculation $
J^\prime(\rho) =  \int
d\phi_1^i d\phi_2^i \delta (\rho - i\phi_1^i\phi_2^i) = \int d\lambda
e^{i\lambda \rho}\int d\phi_1^i
d\phi_2^i~e^{-i\lambda\phi_1^i\phi_2^i} \sim \rho^{-(N+1)} $}
Despite this difference, the final collective Hamiltonian is in fact
{\em identical} to the $O(2N)$ oscillator collective Hamiltonian
\ben
H_{coll}^{Sp(2N)} = -2\frac{\partial}{\partial \rho} \rho  \frac{\partial}{\partial \rho} + \frac{N^2}{2\rho}+\frac{1}{2}k^2\rho 
\label{four-seven}
\een This leads to the same saddle point equation, and the solutions
satisfy the same equation as (\ref{twofive}) with $N \rightarrow 2N$.

In the $O(2N)$ oscillator, we had to choose the positive sign, since
$\sigma$ is by definition a real {\em positive} variable. In this case, there is no reason for $\rho$ to be positive. In fact we need
to choose the negative
sign, since (\ref{four-two1}) requires that the one point function of
$\rho$ must be the negative of the one point function of $\sigma$. 
\ben \rho_0 =
-\frac{N}{k}
\label{four-eight}
\een
It is interesting that the singlet sectors of the $O(2N)$ and $Sp(2N)$ models are described by two different solutions of the {\em same} collective theory.

The leading order ground state energy is the Hamiltonian evaluated on the saddle point,
\ben
E_{gs} = -Nk
\label{cone}
\een
in agreement with (\ref{afour}). 
The fluctuation Hamiltonian is obtained as usual by expanding
\ben
\rho = \rho_0 +\sqrt{\frac{4N}{k}}\xi~~,~~~~~
\Pi_\rho = \sqrt{\frac{k}{4N}}\pi_\xi
\label{four-nine}
\een
The quadratic Hamiltonian is now {\em negative}, essentially because
of the negative sign in the saddle point,
\ben
H^{(2)}_\xi = -\frac{1}{2} \left[ \pi_\xi^2 + 4 k^2 \xi^2 \right]
\label{ctwo}
\een
A standard quantization of this theory leads to a spectrum which is unbounded from below. We will now argue that we need to quantize this theory rather differently, in a way similar to the treatment of 
\cite{shimo}. This involves defining
annihilation and creation operators $a_\xi, a_\xi^\dagger$ 
\ben
\xi = \frac{1}{\sqrt{4k}}[a_\xi + a^\dagger_\xi]~~,~~~~~~\pi_\xi =
i\sqrt{k}[a_\xi - a^\dagger_\xi]
\label{cthree}
\een
which now
satisfy
\ben
[ a_\xi, a_\xi^\dagger ] = -1~~,~~~[ H, a_\xi] = - 2k a_\xi~~,~~~[H,
  a^\dagger_\xi] = 2k a^\dagger_\xi
\label{cfour}
\een
Because of the negative sign of the first commutator in (\ref{cfour}) a standard quantization will lead to a {\em highest energy} state annihilated by $a^\dagger_\xi$, and then the action of powers of $a_\xi$ leads to an infinite tower of states with lower and lower energies. The highest state has a normalizable wavefunction of the standard form $e^{-k\xi^2}$ (Note that the expression for $\pi_\xi$ has a negative sign compared to the usual harmonic oscillator). It is easy to see that this standard quantization does not reproduce the correct two-point function of the $Sp(2N)$ theory, does not lead to the correct spectrum (\ref{afour}) and, as shown below, does not lead to the correct wavefunction.

All this happens because $\rho$ and hence $\xi$ is not really a bosonic variable, and this allows other possibilities.
Consider now a state
$|0\rangle_\xi$ which is annihilated by the annihilation operator $a_\xi$. This leads to a wavefunction ${\rm exp}[k \xi^2]$, which is inadmissible if $\xi$ is really a bosonic variable since it would be non-normalizable. However the true integration is over the Grassmann partons of these collective fields, and in terms of Grassmann integration this wavefunction is perfectly fine.
This is in fact the state which has to be identified
with the ground state of the $Sp(2N)$ oscillator. Including the factor
of the Jacobian, the full wavefunction is (at large $N$)
\ben
\Psi_{0\xi}^\prime[\xi] = [J^\prime(\rho)]^{-1/2}{\rm exp}[k\xi^2]
= [-\frac{N}{k} +2 \sqrt{\frac{N}{k}}\xi]^{N/2}{\rm
  exp}[k\xi^2]
\label{cfive}
\een
Expanding the Jacobian factor in powers of $\xi$ one now sees that the
term which is quadratic in $\xi$ cancels exactly, leaving with
\ben
\Psi_{0\xi}^\prime[\xi] = {\rm exp}[-\sqrt{Nk} \xi + O(\xi^3)]
\label{csix}
\een
This is easily seen to exactly agree with $\Psi_0$ in (\ref{athree})
\ben
\Psi_0 \sim {\rm exp}[-\frac{1}{2}k\rho] \sim {\rm exp}[-\sqrt{Nk}\xi]
\label{cseven}
\een
up to a constant. Once again we need to take into account the
interaction terms in the collective Hamiltonian to  check that the
$O(\xi^3)$ terms cancel.  It can be easily verified that the propagator of fluctuations
$\xi$ will now be {\em negative} of the usual harmonic oscillator
propagator. Furthermore the action of $a^\dagger_\xi$ now generates a tower of states with the energies (\ref{afour}) - except that the integer $n$ is not bounded by $N$.

The fact that we get an unbounded (from above) spectrum from the collective theory is not a surprise. This is an expansion around $N=\infty$ and at $N=\infty$ the spectrum of $Sp(2N)$ is also unbounded.
At finite $N$ a change of variables to $\rho$ is not useful because of the constraints coming from the Grassmann origin of $\rho$.
Nevertheless, even in the $1/N$ expansion, the Grassmann origin allows us to consider wavefunctions which would be otherwise considered inadmissible. 

The negative propagator ensures that the relationship
(\ref{four-two1}) is satisfied for the 2 point functions.
Once this choice is made, the relationship (\ref{four-two1}) holds for
all $m$-point functions to the
leading order in the large-$N$ limit. 
As commented earlier, a term with even
number of $\pi_\xi$ or $\xi$ would have an odd number of factors of
$\rho_0$. Therefore a $n$-point vertex in the theory will differ from
the corresponding $n$-point vertex of the $O(N)$ theory by a factor of
$(-1)^{n+1}$.  
The connected correlator which appears in (\ref{four-two1}) is the sum of
all connected tree diagrams with $n$ external legs. 
The collective theory gives us the following Feynman rules 
\begin{itemize}
    \item [1] Every propagator contributes to a negative sign.
    \item [2] A $p$ point vertex has a factor of $(-1)^{p+1}$
\end{itemize}
We now argue that these rules ensure the validity of the basic relation (\ref{four-two}).
We do it by the following simple diagrammatic method:
\begin{figure}[h]
\centering
\includegraphics[scale=0.50]{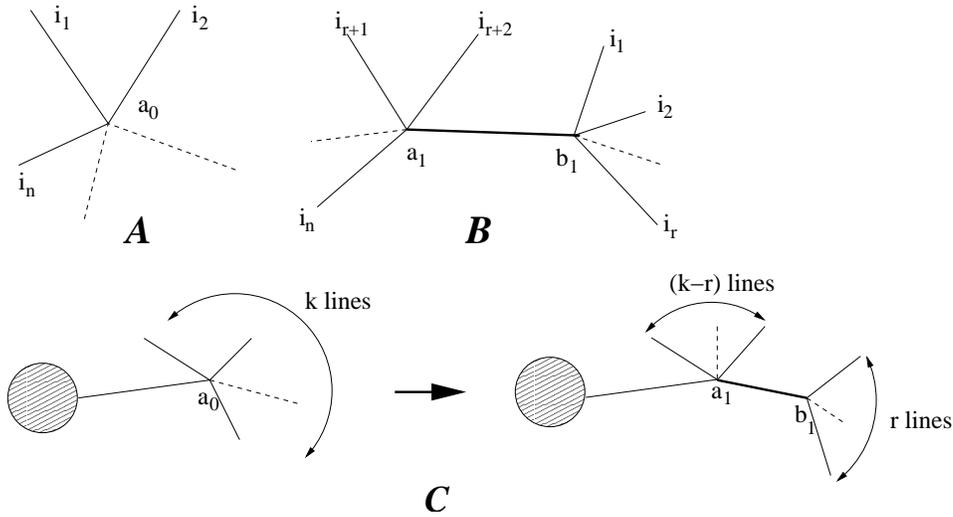}
\caption{Connected tree level correlators of the collective theory}
\label{fig:tree}
\end{figure}

Consider first the simplest diagram for a $n$-point function, figure
A, which is a star graph. The net sign of the diagram is $(-1)^{n+1}
\times (-1)^n = -1$, where the first factor is from the vertex $a_0$
and the second one from the number of lines. Now we proceed to construct all
other tree level diagrams from A, by pulling `$r$' lines resulting in
figure B, which now has vertices, $a_1$ and $b_1$ joined by a new
line. It is easy to see, that the sign of figure A is not changed by
this operation. The net sign of figure B is $(-1)^{(n-r+1)+1} \times
(-1)^{(r+1)+1}\times(-1)^{(n+1)} = -1$, where the 3 factors are from
$a_1$, $b_1$ and the number of lines respectively. In figure C we
repeat this method for the substar diagrams until we exhaust all
possibilities. It is easy to see that the sign stays
invariant. Assigning a sign $\alpha$ to the blob, we first find the
net sign of the left diagram in figure C. It turns out to be, $\alpha
\times (-1)^{(k+1)+1} \times (-1)^{k+1} = - \alpha$. After the
``pulling" operation we get $\alpha \times (-1)^{(k-r+2)+1} \times
(-1)^{(r+1)+1} \times (-1)^{k+1+1} = - \alpha$. Thus it is proved that
in every move the sign is preserved. This proves the relationship
(\ref{four-two1}) for all correlation functions.

\subsection{$Sp(2N)$ Correlators}

Our discussion of the bosonic $O(N)$ collective field theory shows
that the $Sp(2N)$ collective field theory in momentum space is a
straightforward generalization. In this subsection we discuss the
relevant features of the collective theory for the free $Sp(2N)$ model.

The collective Hamiltonian is again exactly the same as in the $O(N)$
theory, given by (\ref{three-three}) with $\sigma \rightarrow
\rho$. Since the connected correlators of the collective fields
satisfy 
\ben 
\langle\rho(\vk_1,\vk_1^\prime, t_1)\rho(\vk_2, \vk_2^\prime
,t_2)\cdots\rho(\vk_n,\vk_n^\prime,t_n)\rangle^{conn}_{Sp(2N)} = -
\langle\sigma(\vk_1,\vk_1^\prime,t_1)\sigma(\vk_2,\vk_2^\prime,t_2) \cdots
\sigma(\vk_n,\vk_n^\prime, t_n)\rangle^{conn}_{SO(2N)}
\label{fifty}
\een
we now need to choose the negative saddle point,
\ben
\rho_0(\vk ,\vk^\prime,t)= -\frac{N}{|\vk|}\delta(\vk -\vk^\prime)
\label{five-one}
\een
The fluctuation Hamiltonian once again has a factor of $(-1)^{n+1}$
for the $n$-point vertex. In particular, the propagator of the
collective field is negative of that of the $O(N)$ collective field -
the quadratic Hamiltonian has an overall negative sign! This is
required - the diagramatic argument for the $Sp(2N)$ oscillator
generalizes in a straightforward fashion, ensuring that (\ref{fifty})
holds.

\section{Bulk Dual of the $Sp(2N)$ model}

In \cite{Das:2003vw}, it was proposed that the collective field theory
for the $d$ dimensional free $O(N)$ theory is in fact Vasiliev's higher spin theory in
AdS$_{d+1}$.  It is easy to see that the collective field has the
right collection of fields.  Consider for example $d=3$. The field
depends on four spatial variables, which may be reorganized as three
spatial coordinates one of which is restricted to be positive and an
angle. A fourier series in the angle then gives rise to a set of
fields $\chi_{\pm n}$ which depend on three spatial variables, with
the integer $n$ denoting the conjugate to the angle. Symmetry under
interchange of the arguments of the collective field then requires $n$
to be even integers. But this is precisely the content of a theory of
massless even spin fields in four space-time dimensions, with $n$
labelling the spin and the two signs corresponding to the two
helicities. (Recall that in four space-time dimensions massless fields
with any spin have just two helicity states).

The precise relationship between collective fields and higher spin
fields in AdS was found in \cite{Koch:2010cy} which we now summarize
for $d=3$. The correspondence is formulated in the  light front
quantization. Denote the usual Minkowski coordinates on the space-time
on which the $O(N)$ fields live by $t,y,x$ and define light cone
coordinates 
\ben 
x^\pm = \frac{1}{\sqrt{2}}(t \pm y)
\label{five-one1}
\een
The conjugate momenta to $x^+,x^-$ are denoted by $p^-,p^+$. Then in
light front quantization where $x^+$ is treated as time, the
Schrodinger picture fields are $\phi^i (x^-,x)$ while the momentum
space fields are given by $\phi^i (p^+,p)$. The corresponding
collective field is then defined as
\ben
\sigma(p^+_1,p_1;p^+_2,p_2)=\phi^i(p^+_1,p_1)\phi^i(p^+_2,p_2)
\label{five-two}
\een The fluctuation of this field around the saddle point is denoted
by $\Psi(p^+_1,p_1;p^+_2,p_2)$. Now define the following bilocal field
\ben 
\Phi(p^+,p^x,z,\theta) = \int dp^z
dp_1^+dp_2^+dp_1dp_2~K(p^+,p^x,z,\theta; p^+_1,p_1,p^+_2,p_2)
\Psi(p^+_1,p_1;p^+_2,p_2)
\label{five-three}
\een
where the kernel is given by
\bea
K(p^+,p^x,z,\theta; p^+_1,p_1,p^+_2,p_2) 
&= & z~e^{izp_z}~ \delta (p^+_1+p^+_2-p^+) ~\delta(p_1+p_2-p) \nn \\ &
& \delta(p_1\sqrt{\frac{p_2^+}{p_1^+}}-
p_2\sqrt{\frac{p_1^+}{p_2^+}}-p^z)~
\delta(2\tan^{-1}\sqrt{\frac{p_2^+}{p_1^+}}-\theta) \nn
\label{five-four}
\eea
In \cite{Koch:2010cy} it was shown that the Fourier transforms of the field
$\Phi(p^+,p^x,z,\theta)$ with respect to $\theta$ satisfy the same linearized equation of motion as the physical helicity modes of 
higher spin gauge fields in AdS$_4$ in light cone gauge. The metric of this AdS$_4$ is given by the standard Poincare form
\ben
ds^2 = \frac{1}{z^2}[-2dx^+dx^-+dx^2+dz^2]=\frac{1}{z^2}[-dt^2+dy^2+dx^2+dz^2]
\label{five-five}
\een
The momenta $p^+,p$ are conjugate to $x^-,x$. The additional dimension generated from the large-N degrees of freedom is $z$, which is canonically conjugate to $p^z$ and is given in terms of the phase space coordinate of the bi-locals by
\ben
z = \frac{(x_1-x_2)\sqrt{p^+_1p^+_2}}{p^+_1+p^+_2}
\label{five-six}
\een
In particular, the linearized equation for the spin zero field, $\varphi (x^-,x,z)$, follows from the quadratic action
\ben
S = \frac{1}{2} \int dx^+dx^-dzdx~\left[\frac{1}{z^2}\left( -2\partial_+\varphi\partial_-\varphi -(\partial_x\varphi)^2-(\partial_z\varphi)^2 \right)+\frac{2}{z^4}\varphi^2 \right]
\label{five-seven1}
\een
which is of course the action of a conformally coupled scalar in the AdS$_4$ with coordinates given by (\ref{five-six}). The actions for the spin-$2s$ fields can be similarly written down. Even though these actions are derived using light cone coordinates, they can be covariantized easily since these are free actions. In terms of the coordinates $t,y,x,z$ the scalar action is given by
\ben
S = \frac{1}{2} \int dt dz dx dy~\left[\frac{1}{z^2}\left( (\partial_t \varphi)^2 -(\partial_y \varphi)^2 -(\partial_x\varphi)^2-(\partial_z\varphi)^2 \right)+\frac{2}{z^4}\varphi^2 \right]
\label{five-seven}
\een

Let us now turn to the $Sp(2N)$ collective theory. One can define once
again the fields as in (\ref{five-three}) and (\ref{five-four}). The
coordinates $(x^+,x^-,x,z)$ will continue to transform appropriately
under AdS isometries. However, we saw earlier that the quadratic
part of the Hamiltonian, and therefore the quadratic part of the
action will have an overall {\em negative} sign.

A negative kinetic term signifies a pathology. Indeed we derived this
theory with the Lorentzian signature $Sp(2N)$ model, which has negative
norm states. The negative kinetic term of the collective theory is
possibly intimately related to this lack of unitarity.

However, the form of the action (\ref{five-seven}) cries out for a {\em analytic continuation}
\ben
z = i\tau~~,~~~~~~t = -iw
\label{five-eight}
\een
Under this continuation the action, $S$  becomes 
\ben
S^\prime = \frac{1}{2} \int d\tau dw dx dy~\left[\frac{1}{\tau^2}\left( (\partial_\tau \varphi)^2 -(\partial_y \varphi)^2 -(\partial_x\varphi)^2-(\partial_w\varphi)^2 \right)-\frac{2}{\tau^4}\varphi^2 \right]
\label{five-nine}
\een
The sign of the mass term {\em has not changed} in this analytic continuation, and this action has become the action of a conformally coupled scalar field in de Sitter space with the metric
\ben
ds^2 = \frac{1}{\tau^2} [ -d\tau^2 + dx^2 + dy^2 + dw^2]
\label{sixty}
\een
This mechanism works for all {\em even} higher spin fields at the
quadratic level. 

To summarize, the collective field theory of the three dimensional
Lorentzian $Sp(2N)$ model can be written as a theory of massless even
spin fields in AdS$_4$, but with negative kinetic terms. Under a
double analytic continuation this becomes the action in dS$_4$ with
positive kinetic terms.  This is consistent with the conjecture of
\cite{Anninos:2011ui} that the {\em euclidean} $Sp(N)$ model is dual
to Vasiliev theory in dS$_4$. It is interesting to note that the way
an emergent holographic direction is similar to the way the Liouville
mode has to be interpeted as a time dimension in worldsheet
supercritical string theory \cite{Das:1988ds}. In this latter case,
the sign of the kinetic term for the Liouville mode is negative for $d
> d_{cr}$.

Even for the $O(N)$ model, the collective field is an represents seemingly an overcomplete description, since for a finite number of points in space $K$, one  replaces at most $NK$ variables by $K^2$ variables, which is much larger in the thermodynamic and continuum limit. However, in the perturbative $1/N$ expansion this is not an issue and the collective theory is known to reproduce the standard results of the $O(N)$ model. The issue becomes of significance at  finite $N$ level.The relevance of incorporating  for such  features  has been noted in \cite{Ng:2012xp,Shenker:2011zf}.

For the fermionic  $Sp(2N)$ model, there appears potentially an even more important redundancy related to the Grassmannian origin of the construction.  Consequently the fields are to obey nontrivial constraint relationships and the Hilbert space is subject to a cutoff of highly excited states. This `exclusion principle'
 was noted already in the AdS correspondence involving $S_N$ orbifolds\cite{Maldacena:1998bw,Jevicki:1998rr,Jevicki:1998bm}.  

In an expansion around $N = \infty$ most effects of this are invisible and our discussion shows that this can be regarded as a theory of higher spin fields in dS is insensitive to these effects. However, as we saw above, the Grassmann origin was already  of importance  in choosing the correct saddle point and the correct quanization of the quadratic hamiltonian. In the next section we will address the question of finite $N$ and the Hilbert space of the bi-local theory. In the framework of  geometric (pseudospin) representation we will give evidence that the bi-local theory is non-perturbatively satisfactory at the finite $N$ level.

\section{Geometric Representation and The Hilbert Space}

The bi-local collective field representation is seen to give a bulk description dS space and the Higher Spin fields. It provides an interacting theory with  vertices governed by $G=1/N$ as the coupling constant.  We would now to show that the collective theory  has an equivalent   geometric (Pseudo) Spin variable description
appropriate for nonperturbative considerations. The essence of this (geometric) description is in reinterpreting  the bi-local collective fields (and their canonical conjugates) as   matrix variables (of infinite dimensionality) endowed with a Kahler structure.This geometric description  will provide a tractable framework for quantization and non-perturbative definition of the bi-local and HS  de Sitter  theory. It will be seen capable  to incorporate non-perturbative features related to the Grassmannian origin of bi-local fields and  its Hilbert space.
Pseudo-spin collective variables  represent all $Sp(2N)$ invariant variables of the theory (both commuting  and non-commuting). These  close  a compact algebra and at large $N$ are constrained by the corresponding Casimir operator. One therefore has an algebraic pseudo-spin system whose nonlinearity is governed by the coupling constant $G=1/N$. As such they have been employed earlier for developing a large $N$ expansion \cite{Jevicki:1980mj}  and  as a model for  quantization \cite{Berezin:1978sn}. This version of the theory is in its perturbative ($1/N$) expansion identical to the bi-local collective representation. It therefore has the same map to and correspondence with Higher Spin  dS$_4$ at perturbative level. We will see however  that the geometric  representation becomes of use for defining
(and evaluating)  the Hilbert space and its quantization.

To describe the pseudo-spin description of the  $Sp(2N)$ theory we will follow the quantization procedure  of \cite{LeClair:2007iy}. In this approach one starts from  the action: 
\begin{equation}
S=\int d^dx\; dt(\partial^\mu\eta^i_1\partial_\mu\eta^i_2)
\end{equation}
and deduces the canonical anti-commutation relations
\begin{equation}
\{\eta^i_1(x,t)\partial_t\eta^j_2(x',t)\}=-\{\eta^i_2(x,t)\partial_t\eta^j_1(x',t)\}=i\delta^d(x-x')\delta^{ij}
\end{equation}
The quantization based on the mode expansion
\begin{eqnarray}
\eta^i_1(x)&=&\int \frac{d^dk}{(2\pi)^{d/2}\sqrt{2\omega_k}}(a_{k+}^{i\dagger} e^{-ikx}+a^i_{k-}e^{ikx})\cr
\eta^i_2(x)&=&\int \frac{d^dk}{(2\pi)^{d/2}\sqrt{2\omega_k}}(-a_{k-}^{i\dagger} e^{-ikx}+a^i_{k+}e^{ikx})
\end{eqnarray}
with 
\begin{eqnarray}
\{a^i_{k-},a_{k'-}^{j\dagger}\}=\{a^i_{k+},a_{k'+}^{j\dagger}\}=\delta^d(k-k')\delta^{ij}
\end{eqnarray}
Note that in this approach the operators $\eta^i_a$ are not hermitian, but pseudo-hermitian in the sense of \cite{Bender:2007nj}.

Pseudo-spin bi-local variables will be introduced based on  $Sp(2N)$ invariance, we have the  vectors:
\begin{eqnarray}
\eta&=&(\eta_1^1,\eta_2^1,\eta_1^2,\eta_2^2,\cdots,\eta_1^N,\eta_2^N)\cr
a(k)&=&(a^1_{k-},a_{k+}^1,a^2_{k-},a_{k+}^2,\cdots ,a^N_{k-},a_{k+}^N)\cr
\tilde{a}(k)&=&(a_{k+}^{1\dagger},-a^{1\dagger}_{k-},a_{k+}^{2\dagger},-a^{2\dagger}_{k-},\cdots ,a_{k\dagger}^{N+},-a^{N\dagger}_{k-})
\end{eqnarray}
and the notation:  
\begin{equation}
\eta(x)=\int \frac{d^dk}{(2\pi)^{d/2}\sqrt{2\omega_k}}(\tilde{a}(k)e^{-ikx}+a(k)e^{ikx})
\end{equation}
so that  a complete set of $Sp(2N)$ invariant operators now follows:
\begin{eqnarray}
S(p_1,p_2)&=&\frac{-i}{2\sqrt{N}}a^T(p_1)\epsilon_N a(p_2)=\frac{i}{2\sqrt{N}}\sum_{i=1}^N(a_{p_1+}^i a_{p_2-}^i+a_{p_2+}^i a_{p_1-}^i)\cr
S^\dagger(p_1,p_2)&=&\frac{-i}{2\sqrt{N}}\tilde{a}^T(p_1)\epsilon_N \tilde{a}(p_2)=\frac{i}{2\sqrt{N}}\sum_{i=1}^N(a_{p_1+}^{i\dagger} a_{p_2-}^{i\dagger}+a_{p_2+}^{i\dagger} a_{p_1-}^{i\dagger})\cr
B(p_1,p_2)&=&\tilde{a}^T(p_1)\epsilon_N a(p_2)=\sum_{i=1}^Na_{p_1+}^{i\dagger}a_{p_2+}^{i}+a_{p_1-}^{i\dagger}a_{p_2-}^{i}
\end{eqnarray}
and  $\epsilon_N=\epsilon \otimes \mathbb{I}_N$, $\epsilon=\left(    \begin{array} {cc} 0&1 \\  -1&0  \end{array}  \right)$

These invariant operators close an invariant algebra. The  commutation relations are found to equal:
\begin{eqnarray}
\big[ S(\vec{p}_1,\vec{p}_2),S^\dagger(\vec{p}_3,\vec{p}_4)\big]=&&{1\over 2}\left(
\delta_{\vec{p}_2,\vec{p}_3}\delta_{\vec{p}_4,\vec{p}_1}+\delta_{\vec{p}_2,\vec{p}_4}\delta_{\vec{p}_3,\vec{p}_1}\right)
-{1\over {4N}}\left[\delta_{\vec{p}_2,\vec{p}_3}B(\vec{p}_4,\vec{p}_1)+\delta_{\vec{p}_2,\vec{p}_4}B(\vec{p}_3,\vec{p}_1)\right.\cr
&&+\left.
\delta_{\vec{p}_1,\vec{p}_3}B(\vec{p}_4,\vec{p}_2)+
\delta_{\vec{p}_1,\vec{p}_4}B(\vec{p}_3,\vec{p}_2)\right]\cr
\big[ B(\vec{p}_1,\vec{p}_2),S^\dagger(\vec{p}_3,\vec{p}_4)\big]=&&
\delta_{\vec{p}_2,\vec{p}_3}S^\dagger(\vec{p}_1,\vec{p}_4)+
\delta_{\vec{p}_2,\vec{p}_4}S^\dagger(\vec{p}_1,\vec{p}_3)\cr
\big[ B(\vec{p}_1,\vec{p}_2),S(\vec{p}_3,\vec{p}_4)\big]=&&-
\delta_{\vec{p}_1,\vec{p}_3}S(\vec{p}_2,\vec{p}_4)-
\delta_{\vec{p}_1,\vec{p}_4}S(\vec{p}_2,\vec{p}_3)
\end{eqnarray}

The singlet sector of the original $Sp(2N)$ theory is characterized by a further constraint. This constraint is is associated with the Casimir operator of of the algebra and can be shown  to take the form:
\begin{eqnarray}
\frac{4}{N}S^\dagger\star S+(1-\frac{1}{N}B)\star (1-\frac{1}{N}B)=\text{$\mathbb{I}$}
\end{eqnarray}
Here we have used the matrix star product  notation: $\star$ product as: with $A\star B=\int d\vec{p}_2A(\vec{p}_1\vec{p}_2)B(\vec{p}_2\vec{p}_3)$. 

The form of the  Casimir, which commutes with the above pseudo-spin fields points to the compact nature of the bi-local pseudo-spin algebra associated with the $Sp(2N)$ theory. This will have major consequences which we will highlight later.

Indeed it is interesting to compare the algebra with the  bosonic case, where we have:
\begin{eqnarray}
S(p_1,p_2)&=&\frac{1}{2\sqrt{N}}\sum_{i=1}^{2N} a_i(p_1)a_i(p_2)\cr
S^\dagger(p_1,p_2)&=&\frac{1}{2\sqrt{N}}\sum_{i=1}^{2N} a_i^\dagger(p_1)a_i^\dagger(p_2)\cr
B(p_1,p_2)&=&\sum_{i=1}^{2N} a_i^\dagger(p_1)a_i(p_2)
\end{eqnarray}
with the  commutation relations:
\begin{eqnarray}
\big[ S(\vec{p}_1,\vec{p}_2),S^\dagger(\vec{p}_3,\vec{p}_4)\big]=&&{1\over 2}\left(
\delta_{\vec{p}_2,\vec{p}_3}\delta_{\vec{p}_4,\vec{p}_1}+\delta_{\vec{p}_2,\vec{p}_4}\delta_{\vec{p}_3,\vec{p}_1}\right)
+{1\over {4N}}\left[\delta_{\vec{p}_2,\vec{p}_3}B(\vec{p}_4,\vec{p}_1)+\delta_{\vec{p}_2,\vec{p}_4}B(\vec{p}_3,\vec{p}_1)\right.\cr
&&+\left.
\delta_{\vec{p}_1,\vec{p}_3}B(\vec{p}_4,\vec{p}_2)+
\delta_{\vec{p}_1,\vec{p}_4}B(\vec{p}_3,\vec{p}_2)\right]\cr
\big[ B(\vec{p}_1,\vec{p}_2),S^\dagger(\vec{p}_3,\vec{p}_4)\big]=&&
\delta_{\vec{p}_2,\vec{p}_3}S^\dagger(\vec{p}_1,\vec{p}_4)+
\delta_{\vec{p}_2,\vec{p}_4}S^\dagger(\vec{p}_1,\vec{p}_3)\cr
\big[ B(\vec{p}_1,\vec{p}_2),S(\vec{p}_3,\vec{p}_4)\big]=&&-
\delta_{\vec{p}_1,\vec{p}_3}S(\vec{p}_2,\vec{p}_4)-
\delta_{\vec{p}_1,\vec{p}_4}S(\vec{p}_2,\vec{p}_3)
\end{eqnarray}
In this case the Casimir constraint is found to equal:
\begin{eqnarray}
-\frac{4}{N}S^\dagger\star S+(1+\frac{1}{N}B)\star (1+\frac{1}{N}B)=\text{$\mathbb{I}$}
\end{eqnarray}
featuring the non-compact nature of the bosonic problem.

We can see therefore that the singlet sectors of the  fermionic $Sp(2N)$ theory and the bosonic $O(2N)$ theory can be described in analogous a bi-local pseudo-spin algebraic formulations with a quadratic Casimir taking the form: 
\begin{eqnarray}
4\gamma S^\dagger\star S+(1-\gamma B)\star (1-\gamma B) =\text{$\mathbb{I}$}
\end{eqnarray}
the difference being that  with $\gamma=\frac{1}{N}(-\frac{1}{N})$ for the fermionic (bosonic) case respectively. This signifies the compact versus the non-compact nature of the algebra, but also exhibits the relationship obtained through the $N \leftrightarrow -N$  switch that was central in the argument for de Sitter correspondence in \cite{Anninos:2011ui}.

From this algebraic bi-local formulation one can easily see the the Collective field representation(s) that we have discussed in sections 2 and 3.  Very simply, the Casimir constraints can be solved, and the algebra implemented in terms of a canonical pair of bi-local fields:
\begin{eqnarray}
S(p_1p_2)&=& \frac{\sqrt{-\gamma}}{2}\int dy_1dy_2e^{-i(p_1y_2+p_2y_2)}\{-\frac{2}{\kappa_{p_1}\kappa_{p_2}}\Pi\star\Psi\star\Pi(y_1y_2)-\frac{1}{2\gamma^2\kappa_{p_1}\kappa_{p_2}}\frac{1}{\Psi}(y_1y_2)\cr
&&+\frac{\kappa_{p_1}\kappa_{p_2}}{2}\Psi(y_1y_2)-i\frac{\kappa_{p_1}}{\kappa_{p_2}}\Psi\star\Pi(y_1y_2)-i\frac{\kappa_{p_2}}{\kappa_{p_1}}\Pi\star\Psi(y_1y_2)\}\cr
S^\dagger(p_1p_2)&=& \frac{\sqrt{-\gamma}}{2}\int dy_1dy_2e^{-i(p_1y_2+p_2y_2)}\{-\frac{2}{\kappa_{p_1}\kappa_{p_2}}\Pi\star\Psi\star\Pi(y_1y_2)-\frac{1}{2\gamma^2\kappa_{p_1}\kappa_{p_2}}\frac{1}{\Psi}(y_1y_2)\cr
&&+\frac{\kappa_{p_1}\kappa_{p_2}}{2}\Psi(y_1y_2)+i\frac{\kappa_{p_1}}{\kappa_{p_2}}\Psi\star\Pi(y_1y_2)+i\frac{\kappa_{p_2}}{\kappa_{p_1}}\Pi\star\Psi(y_1y_2)\}\cr
B(p_1p_2)&=&\frac{1}{\gamma}+\int dy_1dy_2e^{-i(p_1y_2+p_2y_2)}\{\frac{2}{\kappa_{p_1}\kappa_{p_2}}\Pi\star\Psi\star\Pi(y_1y_2)+\frac{1}{2\gamma^2\kappa_{p_1}\kappa_{p_2}}\frac{1}{\Psi}(y_1y_2)\cr
&&+\frac{\kappa_{p_1}\kappa_{p_2}}{2}\Psi(y_1y_2)-i\frac{\kappa_{p_1}}{\kappa_{p_2}}\Psi\star\Pi(y_1y_2)+i\frac{\kappa_{p_2}}{\kappa_{p_1}}\Pi\star\Psi(y_1y_2)\}
\end{eqnarray}
where $\kappa_{p}=\sqrt{\omega_{p}}$.

Recalling that the Hamiltonian is given in terms of  $B$  we now see that its bi-local form is  the same in the fermionic and the bosonic case. This explains the feature that we have established by direct construction in Sec. 2,3. While the bi-local field representation of $B$ is the same in the fermionic and bosonic cases, the difference is seen in the representations of operators $S$ and $S^\dagger$. These operators create singlet states in the Hilbert space and the difference contained in the sign of gamma  implies the opposite shifts for the background fields that we have identified in Sec. 2,3. The algebraic pseudo spin reformulation is therefore seen to account for all the perturbative ($1/N$) features of the the bi-local theory that we have identified in Sec. 2,3. However, in addition and we would like to emphasize that, the algebraic  formulation provides a proper framework  for defining the bi-local   Hilbert space.

\subsection{ Quantization and the Hilbert Space}

The bi-local pseudo-spin algebra has several equivalent representations that turn out to be  useful.  Beside that collective representation that we have explained above, one has the simple oscillator representation:
\begin{eqnarray}
S(p_1,p_2)&=&\alpha \star(1-\frac{1}{N}\alpha^\dagger\star \alpha)^{\frac{1}{2}}(p_1,p_2)\cr
S^\dagger(p_1,p_2)&=&(1-\frac{1}{N}\alpha^\dagger\star \alpha)^{\frac{1}{2}}\star \alpha^\dagger(p_1,p_2)\cr
B(p_1,p_2)&=&2\; \alpha^\dagger\star \alpha(p_1,p_2)
\end{eqnarray}
with standard canonical canonical commutators (or Poisson brackets). 

A more relevant  geometric  representation is obtained through a change:
\begin{eqnarray}
\alpha&=&Z(1+\frac{1}{N}\bar{Z}Z)^{-\frac{1}{2}}\cr
\alpha^\dagger&=&(1+\frac{1}{N}\bar{Z}Z)^{-\frac{1}{2}}\bar{Z}
\end{eqnarray}

The pseudo-spins  in the  $Z$ representation are given by:
\begin{eqnarray}
S(p_1,p_2)&=&Z\star(1+\frac{1}{N}\bar{Z}\star Z)^{-1}(p_1,p_2)\cr
S^\dagger(p_1,p_2)&=&(1+\frac{1}{N}\bar{Z}\star Z)^{-1}\star \bar{Z}(p_1,p_2)\cr
B(p_1,p_2)&=&2\;Z\star (1+\frac{1}{N}\bar{Z} \star Z)^{-1}\star \bar{Z}(p_1,p_2)
\end{eqnarray}

It's easy to see that this satisfy the Casimir constraint: $\frac{4}{N}S^\dagger\star S+(1-\frac{1}{N}B)^2=1$

One can write the Lagrangian in this  $Z$ representation as:
\begin{eqnarray}
\text{$\mathcal{L}$}=i\int dt \;\text{tr}[Z(1+\frac{1}{N}\bar{Z}Z)^{-1}\dot{\bar{Z}}-\dot{Z}(1+\frac{1}{N}\bar{Z}Z)^{-1}\bar{Z}]-\text{$\mathcal{H}$}
\end{eqnarray}

For regularization purposes, it is useful to consider putting $\vec{x}$ in a box and limiting the momenta by a cutoff $\Lambda$: this makes the bi-local fields into finite dimensional matrices (which we will take to be a size $K$). For $Sp(2N)$ one deals with a $K\times K$ dimensional complex matrix $Z$ and we have obtained in the above a compact symmetric (Kahler) space :

\begin{equation}
ds^2=\text{tr}[dZ(1-\bar{Z}Z)^{-1}d\bar{Z}(1-Z\bar{Z})^{-1}]
\end{equation}

According to the classification of \cite{Berezin:1975}, this would correspond to manifold $M_I(K,K)$. We note that the standard fermionic problem which was considered in detail in \cite{Berezin:1978sn} corresponds to manifold $M_{\MakeUppercase{\romannumeral 3}}(K,K)$ of complex antisymmetric matrices. 
 
 Quantization on Kahler manifolds in general has been formulated  in detail by Berezin \cite{Berezin:1978sn}. We also note that the usefullnes of Kahler quantization for discretizing de Sitter space was pointed out by  A. Volovich in a quantum mechanical scenario\cite{Volovich:2001rt}.
 In the present Quantization we are dealing with a field theory with infinitely many degrees of freedom and infinite Khaler matrix variables.
 We will now summarize some of the results of quantization which are directly relevant  to the $Sp(2N)$ bi-local collective fields theory. Commutation relations of this system follow from the Poisson Brackets associated with the Lagrangian $\mathcal{L}(\bar{Z},Z)$. States in the Hilbert space are represented  by (holomorphic) functions (functionals) of the bi-locals $Z(k,l)$. A Kahler scalar product defining the bi-local Hilbert space reads:
\begin{equation}
(F_1,F_2)=C(N,K)\int d\mu(\bar{Z},Z)F_1(Z)F_2(\bar{Z})\det [1+\bar{Z}Z]^{-N}
\end{equation}
with the (Kahler) integration measure:
\begin{eqnarray}
d\mu=\det[1+\bar{Z}Z]^{-2K}d\bar{Z}dZ
\end{eqnarray}

The normalization constant is found from requiring $(F_1,F_1)=1$ for $F=1$. Let:
\begin{equation}
a(N,K)=\frac{1}{C(N,K)}=\int d\mu(\bar{Z},Z)\det [1+\bar{Z}Z]^{-N}
\end{equation}

This leads to the matrix integral (complex Penner Model) 
\begin{equation}
a(N,K)=\frac{1}{C(N,K)}=\int \prod_{k,l=1}^K d\bar{Z}(k,l)dZ(k,l)\det [1+\bar{Z}Z]^{-2K-N}
\end{equation}
which determines $C(N,K)$.

The following results on quantization of this type of Kahler system are of note: First, the parameter $N$: much like for ordinary spin, one can show that $N$ (and therefore $G$ in Higher Spin Theory) can only take integer values, i.e. $N=0,1,2,3,\cdots$. Next, one has question about the total  number of states in the above Hilbert space. Naively, bi-local theory would seem to grossly overcount the number of states of the original fermionic theory. Originally one essentially had $2NK$ fermionic degrees of freedom with a finite Hilbert space. The bi-local description is based on (complex) bosonic variables of dimensions $K^2$ and the corresponding Hilbert space would appear to be  much larger. But due to the compact nature of the phase space, the number of states much smaller. 

We will now evaluate this number (at finite $N$ and $K$) for the present case of $Sp(2N)$ (in \cite{Berezin:1978sn} ordinary fermions were studied) and show that the exact  dimension of the bi-local Hilbert space in geometric (Kahler) quantization agrees with the dimension of the singlet Hilbert space of the $Sp(2N)$ fermionic theory. 

The dimension of quantized Hilbert space is found as follows: Considering the operator $\hat{O}=I$ one has that:
\begin{equation}
\text{Tr}(I)=C(N,K)\int \prod_{k,l=1}^K d\bar{Z}(k,l)dZ(k,l)\det [1+\bar{Z}Z]^{-2K}
\end{equation}
Consequently the dimension of the bi-local Hilbert space is given by:
\begin{equation}
\text{Dim  $\mathcal{H}$}_B=\frac{C(N,K)}{C(0,K)}=\frac{a(0,K)}{a(N,K)}
\end{equation}

The evaluation of the matrix (Penner) integral therefore also determines the dimension of the bi-local Hilbert space. Since this evaluation is a little bit involved, we present it in the following. Evaluation of matrix integrals (for real matrices) is given in \cite{Brezin:1977sv} the extension to the complex case was considered 
in\cite{Ginibre:1965}.

We will use results of \cite{Berezin:1975}, whereby every (complex) matrix can be reduced through (symmetry) transformations to a diagonal form:
\begin{eqnarray}
Z(k,l)\rightarrow \begin{bmatrix}
\omega_1 & & & & \\
& \omega_2 & &  \text{\huge{0}} & \\
& & \omega_3 &  & \\
&  \text{\huge{0}} & & \ddots &\\
& & & & \omega_K
\end{bmatrix}
\end{eqnarray}

and the matrix integration measure becomes:
\begin{eqnarray}
[d\bar{Z}dZ]=\vert \Delta(\omega) \vert^2 \prod_{l=1}^K d\omega_l d\Omega
\end{eqnarray}

where $d\Omega$ denotes ``angular" parts of the integration and $\Delta(x_1,\cdots,x_K)=\prod_{k<l}(x_k-x_l)$ is a Vandermonde determinant, with $x_i=\omega_i^2$. Consequently the matrix integral for $a(N,K)$ (and $C(N,K)$) becomes:

\begin{eqnarray}
a(N,K)=\frac{\text{Vol} \;\;\Omega}{K!}\int \Delta(x_1,\cdots,x_K)^2 \prod_l (1+\omega_l^2)^{-2K-N}\prod_l d\omega_l
\end{eqnarray}

changing variables:  $x_i=-\frac{y_i}{1-y_i}$, we get:
\begin{eqnarray}
a(N,K)=\frac{\text{Vol} \;\;\Omega}{2^K K!}\int_0^\Lambda \prod_i^K dy_i \Delta(y_1,\cdots,y_K)^2 \prod_i (1-y_i)^{N}
\end{eqnarray}

This integral can be evaluated exactly. It belongs to a class of integrals evaluated by Selberg in 1944 \cite{Selberg:1944}:
\begin{eqnarray}
I(\alpha,\beta,\gamma,n)&=&\int_0^1 dx_1\cdots \int_0^1dx_n \vert\Delta(x)\vert^{2\gamma} \prod_{j=1}^n x_j^{\alpha-1}(1-x_j)^{\beta-1}\cr
&=&\prod_{j=0}^{n-1}\frac{\Gamma(1+\gamma+j\gamma)\Gamma(\alpha+j\gamma)\Gamma(\beta+j\gamma)}{\Gamma(1+\gamma)\Gamma(\alpha+\beta+(n+j-1)\gamma)}
\end{eqnarray}

we have the case with $\alpha=1,\;\;\beta=N+1,\;\;\gamma=1,\;\;n=K$ and 
\begin{eqnarray}
I(1,N+1,1,K)&=&\prod_{j=0}^{K-1}\frac{\Gamma(2+j)\Gamma(1+j)\Gamma(N+1+j)}{\Gamma(2)\Gamma(N+K+j+1)}
\end{eqnarray}

We therefore obtain the following formula for the number of states in our Bi-local $Sp(2N)$ Hilbert space:

\begin{equation}
\text{Dim  $\mathcal{H}$}_B=\prod_{j=0}^{K-1}\frac{\Gamma(j+1)\Gamma(N+K+j+1)}{\Gamma(K+j+1)\Gamma(N+j+1)} \label{counting}
\end{equation}

We have compared this number with explicit enumeration of $Sp(2N)$ invariant states in the fermionic Hilbert space (for low values of $N$ and $K$) and found complete agreement. It is probably not that difficult to prove agreement for all $N,K$. 
This settles however the potential problem of overcompletness of the bi-local representation. Since the $Sp(2N)$ counting uses the fermionic nature of creation operators and features exclusion when occupation numbers grow above certain limit it is seen that bi-local geometric quantization elegantly incorporates these effects. The compact nature of the associated infinite dimensional Kahler manifold secures the correct dimensionality of the the singlet  Hilbert space. By using Stirling's approximation for the number of states in the bi-local Hilbert space \eqref{counting}, we see the dimension growing linearly in $N$ (with $K \gg N$):
\begin{equation}
\ln (\text{Dim  $\mathcal{H}$}_B)\sim 2NK\;\ln 2     \qquad\text{at the leading order}
\end{equation}
This is a clear demonstration of the presence of an $N$-dependent cutoff in agreement with the fermionic nature of the original $Sp(2N)$ Hilbert space. So in the nonlinear bi-local theory with $G=1/N$ as coupling constant, we have the desired effect that the Hilbert space is cutoff through $1/G$ effects. Consequently we conclude  that the geometric bi-local representation  with infinite dimensional matrices $Z (k,l)$  provides a complete framework for quantization of the bi-local theory and  of de Sitter HS Gravity.

\section{Comments} 
We have motivated the use of double analytic continuation and hence the connection between the $Sp(2N)$ model and de Sitter higher field theory for the quadratic action for the collective field. To establish this connection one of course needs to establish this for the interaction terms. This is of course highly nontrivial, and in fact the connection between the collective theory for the $O(N)$ model and the AdS higher spin theory is only beginning to be understood. We believe that once this is understood well enough one can address the question for the $Sp(2N)$-dS connection.

In this paper we have dealt mostly with the {\em  free} $Sp(2N)$ vector model. As the parallel  $O(N)$/AdS case this theory is characterized  with  an infinite sequence of conserved higher spin currents and associated conserved charges. The question regarding the  implementation of the Coleman-Mandula theorem then arises, this question was discussed recently in \cite{Maldacena:2011jn,Maldacena:2012sf,Koch:2012vc}. One can expected that identical conclusions hold for the present $Sp(2N)$ case.
 The bi-local collective field theory technqiue is trivially extendible to the linear sigma model based on $Sp(2N)$, as commented in section (4.2). Of particular interest is the IR behavior of the theory which presumably takes the theory from the Gaussian fixed point to a nontrivial fixed point. 

It is well known that dS/CFT correspondence is quite different from AdS/CFT correspondence, particularly in the interpretation of bulk correlation functions \cite{dS, dS2}. We have not addressed these  issues in this paper. Recently it has been proposed that the $Sp(2N)$/dS connection can be used to understand subtle points about dS/CFT \cite{Ng:2012xp}. We hope that an explicit construction as described in this paper will be valuable for a deeper understanding of these issues.

The bi-local formulation  that we have presented was  cast in a geometric, pseudo-spin framework. We have suggested that this representation  offers the best framework for quantization of the bi-local  theory and consequently  the Hilbert space in dS/CFT. We have demonstrated through counting of the size  of the
Hilbert space that it incorporates finite $N$ effects through  a cutoff  which depends on the coupling constant of the theory: $G=1/N$. Most importantly it incorporates the finite $N$ exclusion principle and provides an explanation on the quantization of $G=1/N$ from the bulk point of view. These features are obviously of definite relevance for understanding  quantization of Gravity in de Sitter space-time. Nevertheless the question of understanding de Sitter Entropy from this   3 dimensional CFT remains an interesting and challenging problem.

It would be interesting  to consider the analogues of $Sp(2N)$/dS correspondence in the  CFT$_2$/Chern-Simons version\cite{Gaberdiel:2010pz,Henneaux:2010xg,Campoleoni:2010zq}, as well as to three dimensional conformal theories which have a line of fixed points, as in \cite{Giombi:2011kc}. Finally higher spin theories arise as limits of string theory in several contexts, e.g. \cite{Sezgin:2002rt} and \cite{Giombi:2011kc}. It would be interesting to see if these models can be modified to realize a dS/CFT correspondence in string theory.

\section{Acknowledgements}

The work of D.D. and S.R.D. is partially supported by National Science Foundation grants PHY-0970069 and PHY-0855614. The work of A.J. and Q.Y. is supported by the Department of Energy under contract DE-FG-02-91ER40688. We would like to thank  Matthew Dodelson, Misha Eides, Igor Klebanov, Gautam Mandal,  Shinji Mukhoyama, Marcus Spradlin, Alfred Shapere and Anastasia Volovich for discussions and Luis Alvarez-Gaume for a comment. S.R.D. thanks PCTS at Princeton University, Theory Group at CERN and University of Barcelona for hospitality during the final stages of the preparation of this manuscript.


\begin{thebibliography}{10}

\bibitem{Klebanov:2002ja} 
  I.~R.~Klebanov and A.~M.~Polyakov,
  Phys.\ Lett.\ B {\bf 550}, 213 (2002)
  [hep-th/0210114].

\bibitem{Vasiliev:2012vf} 
  For a recent review and a more complete list of original references , see 
M.~A.~Vasiliev,
  arXiv:1203.5554 [hep-th].

\bibitem{Giombi:2009wh} 
  S.~Giombi and X.~Yin,
  JHEP {\bf 1009}, 115 (2010)
  [arXiv:0912.3462 [hep-th]].

\bibitem{Sezgin:2003pt} 
  E.~Sezgin and P.~Sundell,
  JHEP {\bf 0507}, 044 (2005)
  [hep-th/0305040].


\bibitem{Jevicki:1979mb} 
  A.~Jevicki and B.~Sakita,
  Nucl.\ Phys.\ B {\bf 165}, 511 (1980).


\bibitem{Das:2003vw}
 S.~R.~Das and A.~Jevicki,
 ``Large-N collective fields and holography,''
 Phys.\ Rev.\  D {\bf 68}, 044011 (2003)
 [arXiv:hep-th/0304093].


\bibitem{Koch:2010cy} 
  R.~d.~M.~Koch, A.~Jevicki, K.~Jin and J.~P.~Rodrigues,
  Phys.\ Rev.\ D {\bf 83}, 025006 (2011)
  [arXiv:1008.0633 [hep-th]].

\bibitem{Jevicki:2011aa} 
  A.~Jevicki, K.~Jin and Q.~Ye,
  arXiv:1112.2656 [hep-th].


\bibitem{Jevicki:2011ss} 
  A.~Jevicki, K.~Jin and Q.~Ye,
  J.\ Phys.\ A A {\bf 44}, 465402 (2011)
  [arXiv:1106.3983 [hep-th]].

\bibitem{Douglas:2010rc} 
  M.~R.~Douglas, L.~Mazzucato and S.~S.~Razamat,
  Phys.\ Rev.\ D {\bf 83}, 071701 (2011)
  [arXiv:1011.4926 [hep-th]].
  
  
\bibitem{Das:1990kaa} 
  S.~R.~Das and A.~Jevicki,
  Mod.\ Phys.\ Lett.\ A {\bf 5}, 1639 (1990).
 

\bibitem{Eguchi:1982nm} 
  T.~Eguchi and H.~Kawai,
  Phys.\ Rev.\ Lett.\  {\bf 48}, 1063 (1982).

\bibitem{Banks:1996vh} 
  T.~Banks, W.~Fischler, S.~H.~Shenker and L.~Susskind,
  Phys.\ Rev.\ D {\bf 55}, 5112 (1997)
  [hep-th/9610043].


\bibitem{Ishibashi:1996xs} 
  N.~Ishibashi, H.~Kawai, Y.~Kitazawa and A.~Tsuchiya,
  Nucl.\ Phys.\ B {\bf 498}, 467 (1997)
  [hep-th/9612115].

\bibitem{dS}
A.~Strominger,
  JHEP {\bf 0110}, 034 (2001)
  [hep-th/0106113];
A.~Strominger,
  JHEP {\bf 0111}, 049 (2001)
  [hep-th/0110087];

\bibitem{Anninos:2011ui} 
  D.~Anninos, T.~Hartman and A.~Strominger,
  arXiv:1108.5735 [hep-th].

\bibitem{Das:1988ds} 
  S.~R.~Das, S.~Naik and S.~R.~Wadia,
  Mod.\ Phys.\ Lett.\ A {\bf 4}, 1033 (1989).

\bibitem{Ng:2012xp} 
  G.~S.~Ng and A.~Strominger,
  arXiv:1204.1057 [hep-th].


\bibitem{witten}
E.~Witten,
  hep-th/0106109;

\bibitem{Balasubramanian:2001rb} 
  V.~Balasubramanian, P.~Horava and D.~Minic,
  JHEP {\bf 0105}, 043 (2001)
  [hep-th/0103171].


  \bibitem{Banks:2003cg} 
  T.~Banks,
  astro-ph/0305037.


\bibitem{Volovich:2001rt} 
  A.~Volovich,
  hep-th/0101176.


\bibitem{Parikh:2004wh} 
  M.~K.~Parikh and E.~P.~Verlinde,
  JHEP {\bf 0501}, 054 (2005)
  [hep-th/0410227].
  


  
\bibitem{Lowe:2004nw} 
  D.~A.~Lowe,
  Phys.\ Rev.\ D {\bf 70}, 104002 (2004)
  [hep-th/0407188].


\bibitem{Dyson:2002nt} 
  L.~Dyson, J.~Lindesay and L.~Susskind,
  JHEP {\bf 0208}, 045 (2002)
  [hep-th/0202163];
N.~Goheer, M.~Kleban and L.~Susskind,
  JHEP {\bf 0307}, 056 (2003)
  [hep-th/0212209].

\bibitem{Henneaux:1982ma} 
  M.~Henneaux and C.~Teitelboim,
  Annals Phys.\  {\bf 143}, 127 (1982);
  M.~Henneaux and C.~Teitelboim,
  Princeton, USA: Univ. Pr. (1992) 520 p

\bibitem{Finkelstein:1985cx} 
  R.~Finkelstein and M.~Villasante,
  Phys.\ Rev.\ D {\bf 33}, 1666 (1986).


\bibitem{deMelloKoch:1996mj} 
  R.~de Mello Koch and J.~P.~Rodrigues,
  Phys.\ Rev.\ D {\bf 54}, 7794 (1996)
  [hep-th/9605079].


\bibitem{Jackiw:1980xj} 
  R.~Jackiw and A.~Strominger,
  Phys.\ Lett.\ B {\bf 99}, 133 (1981).


\bibitem{shimo} H.~Shimodaira, Nucl. Phys. {\bf 17},486 (1960).

\bibitem{Shenker:2011zf} 
  S.~H.~Shenker and X.~Yin,
  arXiv:1109.3519 [hep-th].

\bibitem{Maldacena:1998bw} 
  J.~M.~Maldacena and A.~Strominger,
  JHEP {\bf 9812}, 005 (1998)
  [hep-th/9804085].

\bibitem{Jevicki:1998rr} 
  A.~Jevicki and S.~Ramgoolam,
  JHEP {\bf 9904}, 032 (1999)
  [hep-th/9902059].

\bibitem{Jevicki:1998bm} 
  A.~Jevicki, M.~Mihailescu and S.~Ramgoolam,
  Nucl.\ Phys.\ B {\bf 577}, 47 (2000)
  [hep-th/9907144].

\bibitem{Jevicki:1980mj} 
  A.~Jevicki and N.~Papanicolaou,
  Nucl.\ Phys.\ B {\bf 171}, 362 (1980).


\bibitem{Berezin:1978sn} 
  F.~A.~Berezin,
  Commun.\ Math.\ Phys.\  {\bf 63}, 131 (1978).

\bibitem{LeClair:2007iy} 
  A.~LeClair and M.~Neubert,
  JHEP {\bf 0710}, 027 (2007)
  [arXiv:0705.4657 [hep-th]].

\bibitem{Bender:2007nj} 
  C.~M.~Bender,
  Rept.\ Prog.\ Phys.\  {\bf 70}, 947 (2007)
  [hep-th/0703096 [HEP-TH]].

\bibitem{Berezin:1975}
F.A. Berezin,
{\em Quantization in complex symmetric spaces},
Math. USSR-Izv. {\bf   9} (1975), 341--379.


\bibitem{Brezin:1977sv} 
  E.~Brezin, C.~Itzykson, G.~Parisi and J.~B.~Zuber,
  Commun.\ Math.\ Phys.\  {\bf 59}, 35 (1978).
\bibitem{Ginibre:1965} J. Ginibre, J. Math. Phys. {\bf 6} (1965) 440.


\bibitem{Selberg:1944} 
  A.~Selberg,
  Norsk  Mat.\ Tidsskr.\  {\bf 26}, 71 (1944).

\bibitem{Maldacena:2011jn} 
  J.~Maldacena and A.~Zhiboedov,
  ``Constraining Conformal Field Theories with A Higher Spin Symmetry,''  
  arXiv:1112.1016 [hep-th].  


\bibitem{Maldacena:2012sf} 
  J.~Maldacena and A.~Zhiboedov,
  ``Constraining conformal field theories with a slightly broken higher spin symmetry,''  
  arXiv:1204.3882 [hep-th].  


\bibitem{Koch:2012vc} 
  R.~d.~M.~Koch, A.~Jevicki, K.~Jin, J.~P.~Rodrigues and Q.~Ye,
  arXiv:1205.4117 [hep-th].


\bibitem{dS2}
J.~M.~Maldacena,
  JHEP {\bf 0305}, 013 (2003)
  [astro-ph/0210603];
D.~Harlow and D.~Stanford,
  arXiv:1104.2621 [hep-th].



\bibitem{Giombi:2010vg} 
  S.~Giombi and X.~Yin,
  ``Higher Spins in AdS and Twistorial Holography,''  
  JHEP {\bf 1104}, 086 (2011)  
  [arXiv:1004.3736 [hep-th]].  

\bibitem{Henneaux:2010xg} 
  M.~Henneaux and S.~J.~Rey,
  ``Nonlinear $W_{\infty}$ as Asymptotic Symmetry of Three-Dimensional Higher Spin Anti-de Sitter Gravity,''  
  JHEP {\bf 1012}, 007 (2010)  
  [arXiv:1008.4579 [hep-th]].  

\bibitem{Campoleoni:2010zq} 
  A.~Campoleoni, S.~Fredenhagen, S.~Pfenninger and S.~Theisen,
  ``Asymptotic symmetries of three-dimensional gravity coupled to higher-spin fields,''  
  JHEP {\bf 1011}, 007 (2010)  
  [arXiv:1008.4744 [hep-th]].  


\bibitem{Gaberdiel:2010pz} 
  M.~R.~Gaberdiel and R.~Gopakumar,
  ``An AdS$_3$ Dual for Minimal Model CFTs,''  
  Phys.\ Rev.\ D {\bf 83}, 066007 (2011)  
  [arXiv:1011.2986 [hep-th]].  
  
  
\bibitem{Giombi:2011kc} 
  S.~Giombi, S.~Minwalla, S.~Prakash, S.~P.~Trivedi, S.~R.~Wadia and X.~Yin,
  arXiv:1110.4386 [hep-th].


\bibitem{Sezgin:2002rt} 
  E.~Sezgin and P.~Sundell,
  ``Massless higher spins and holography,''  
  Nucl.\ Phys.\ B {\bf 644}, 303 (2002)  
  [Erratum-ibid.\ B {\bf 660}, 403 (2003)]  
  [hep-th/0205131].  

  


\end{thebibliography}
\end{document}